
\magnification=\magstep1


\xdef \FourScal {\char 65}
\xdef \PhotA {\char 66}
\xdef \PhotB {\char 67}
\xdef \PhotC {\char 68}
\xdef \PhotH {\char 69}

\newdimen\fullhsize
\newdimen\hstitle
\hstitle=\hsize 
\newdimen\hsbody
\hsbody=\hsize 
\newdimen\hbodyoffset
\hbodyoffset=\hoffset 
\newbox\leftpage
\def\abstract#1{#1}
\def\rotated{\special{ps: landscape}
\magnification=1000  
\baselineskip=14pt
\global\hstitle=9truein\global\hsbody=4.75truein
\global\vsize=7truein\global\voffset=-.31truein
\global\hoffset=-0.54in\global\hbodyoffset=-.54truein
\global\fullhsize=10truein
\def\DefineTeXgraphics{%
\special{ps::[global]
/TeXgraphics {currentpoint translate 0.7 0.7 scale
              -80 0.72 mul -1000 0.72 mul translate} def}}
\let\lr=L
\def\ifsmall{\iftrue}
\def\titlepagefont{\twelvepoint}
\trueseventeenpoint
\def\almostshipout##1{\if L\lr \count1=1
      \global\setbox\leftpage=##1 \global\let\lr=R
   \else \count1=2
      \shipout\vbox{\hbox to\fullhsize{\box\leftpage\hfil##1}}
      \global\let\lr=L\fi}

\output={\ifnum\count0=1 
 \shipout\vbox{\hbox to \fullhsize{\hfill\pagebody\hfill}}\advancepageno
 \else
 \almostshipout{\leftline{\vbox{\pagebody\makefootline}}}\advancepageno
 \fi}

\def\abstract##1{{\leftskip=1.5in\rightskip=1.5in ##1\par}} }

\def\linemessage#1{\immediate\write16{#1}}

\global\newcount\secno \global\secno=0
\global\newcount\appno \global\appno=0
\global\newcount\meqno \global\meqno=1
\global\newcount\subsecno \global\subsecno=0
\global\newcount\figno \global\figno=0
\global\newcount\mfino \global\mfino=65

\newif\ifAnyCounterChanged
\let\terminator=\relax
\def\normalize#1{\ifx#1\terminator\let\next=\relax\else%
\if#1i\aftergroup i\else\if#1v\aftergroup v\else\if#1x\aftergroup x%
\else\if#1l\aftergroup l\else\if#1c\aftergroup c\else%
\if#1m\aftergroup m\else%
\if#1I\aftergroup I\else\if#1V\aftergroup V\else\if#1X\aftergroup X%
\else\if#1L\aftergroup L\else\if#1C\aftergroup C\else%
\if#1M\aftergroup M\else\aftergroup#1\fi\fi\fi\fi\fi\fi\fi\fi\fi\fi\fi\fi%
\let\next=\normalize\fi%
\next}
\def\makeNormal#1#2{\def\doNormalDef{\edef#1}\begingroup%
\aftergroup\doNormalDef\aftergroup{\normalize#2\terminator\aftergroup}%
\endgroup}

\def\warnIfChanged#1#2{%
\ifundef#1
\else\begingroup%
\edef\oldDefinitionOfCounter{#1}\edef\newDefinitionOfCounter{#2}%
\ifx\oldDefinitionOfCounter\newDefinitionOfCounter%
\else%
\linemessage{Warning: definition of \noexpand#1 has changed.}%
\global\AnyCounterChangedtrue\fi\endgroup\fi}

\def\PageNo#1{\xdef#1{\the\count0}%
\ifWritingAuxFile\immediate\write\auxfile{\noexpand\xdef\noexpand#1{#1}}\fi%
\noindent\ignorespaces}

\def\Section#1{\xdef\names{#1}
\global\titletrue
\global\advance\secno by1\relax\global\meqno=1%
\global\mfino=65%
\global\subsecno=0%
\bigbreak\bigskip
\centerline{\fourteenbf %
\the\secno. #1}%
\par\nobreak\medskip\nobreak}
\def\tagsection#1{%
\warnIfChanged#1{\the\secno}%
\xdef#1{\the\secno}%
\ifWritingAuxFile\immediate\write\auxfile{\noexpand\xdef\noexpand#1{#1}}\fi%
}
\def\section{\Section}
\def\Subsection#1{\global\advance\subsecno by1\relax\bigskip %
\leftline{\bf\the\secno.\the\subsecno\ #1}%
\par\nobreak\smallskip\nobreak}
\def\tagsubsection#1{%
\warnIfChanged#1{\the\secno.\the\subsecno}%
\xdef#1{\the\secno.\the\subsecno}%
\ifWritingAuxFile\immediate\write\auxfile{\noexpand\xdef\noexpand#1{#1}}\fi%
}

\def\subsection{\Subsection}

\def\romappno{\uppercase\expandafter{\romannumeral\appno}}
\def\makeNormalizedRomappno{%
\expandafter\makeNormal\expandafter\normalizedromappno%
\expandafter{\romannumeral\appno}%
\edef\normalizedromappno{\uppercase{\normalizedromappno}}}
\def\Appendix#1{\xdef\names{#1}
\global\titletrue
\global\advance\appno by1\relax\global\meqno=1\global\secno=0%
\global\mfino=65%
\global\subsecno=0%
\bigbreak\bigskip
\centerline{\twelvepoint \bf Appendix %
\romappno. #1}%
\par\nobreak\bigskip\nobreak}
\def\tagappendix#1{\makeNormalizedRomappno%
\warnIfChanged#1{\normalizedromappno}%
\xdef#1{\normalizedromappno}%
\ifWritingAuxFile\immediate\write\auxfile{\noexpand\xdef\noexpand#1{#1}}\fi%
}
\def\appendix{\Appendix}
\def\Subappendix#1{\global\advance\subsecno by1\relax\bigskip %
\leftline{\bf\romappno.\the\subsecno\ #1}%
\par\nobreak\smallskip\nobreak}
\def\tagsubappendix#1{\makeNormalizedRomappno%
\warnIfChanged#1{\normalizedromappno.\the\subsecno}%
\xdef#1{\normalizedromappno.\the\subsecno}%
\ifWritingAuxFile\immediate\write\auxfile{\noexpand\xdef\noexpand#1{#1}}\fi%
}

\def\nolabels{\def\wrlabeL##1{}\def\eqlabeL##1{}\def\reflabeL##1{}}
\def\writelabels{\def\wrlabeL##1{\leavevmode\vadjust{\rlap{\smash%
{\line{{\escapechar=` \hfill\rlap{\sevenrm\hskip.03in\string##1}}}}}}}%
\def\eqlabeL##1{{\escapechar-1\rlap{\sevenrm\hskip.05in\string##1}}}%
\def\reflabeL##1{\noexpand\rlap{\noexpand\sevenrm[\string##1]}}}
\def\writeleftlabels{\def\wrlabeL##1{\leavevmode\vadjust{\rlap{\smash%
{\line{{\escapechar=` \hfill\rlap{\sevenrm\hskip.03in\string##1}}}}}}}%
\def\eqlabeL##1{{\escapechar-1%
\rlap{\sixrm\hskip.05in\string##1}%
\llap{\sevenrm\string##1\hskip.03in\hbox to \hsize{}}}}%
\def\reflabeL##1{\noexpand\rlap{\noexpand\sevenrm[\string##1]}}}
\nolabels

\def\eqn#1{\makeNormalizedRomappno%
\ifnum\secno>0%
  \warnIfChanged#1{\the\secno.\the\meqno}%
  \eqno(\the\secno.\the\meqno)\xdef#1{\the\secno.\the\meqno}%
     \global\advance\meqno by1
\else\ifnum\appno>0%
  \warnIfChanged#1{\normalizedromappno.\the\meqno}%
  \eqno({\rm\romappno}.\the\meqno)%
      \xdef#1{\normalizedromappno.\the\meqno}%
     \global\advance\meqno by1
\else%
  \warnIfChanged#1{\the\meqno}%
  \eqno(\the\meqno)\xdef#1{\the\meqno}%
     \global\advance\meqno by1
\fi\fi%
\eqlabeL#1%
\ifWritingAuxFile\immediate\write\auxfile{\noexpand\xdef\noexpand#1{#1}}\fi%
}
\def\defeqn#1{\makeNormalizedRomappno%
\ifnum\secno>0%
  \warnIfChanged#1{\the\secno.\the\meqno}%
  \xdef#1{\the\secno.\the\meqno}%
     \global\advance\meqno by1
\else\ifnum\appno>0%
  \warnIfChanged#1{\normalizedromappno.\the\meqno}%
  \xdef#1{\normalizedromappno.\the\meqno}%
     \global\advance\meqno by1
\else%
  \warnIfChanged#1{\the\meqno}%
  \xdef#1{\the\meqno}%
     \global\advance\meqno by1
\fi\fi%
\eqlabeL#1%
\ifWritingAuxFile\immediate\write\auxfile{\noexpand\xdef\noexpand#1{#1}}\fi%
}
\def\anoneqn{\makeNormalizedRomappno%
\ifnum\secno>0
  \eqno(\the\secno.\the\meqno)%
     \global\advance\meqno by1
\else\ifnum\appno>0
  \eqno({\rm\normalizedromappno}.\the\meqno)%
     \global\advance\meqno by1
\else
  \eqno(\the\meqno)%
     \global\advance\meqno by1
\fi\fi%
}
\def\mfig#1#2{\global\advance\figno by1%
\relax#1\the\figno%
\warnIfChanged#2{\the\figno}%
\edef#2{\the\figno}%
\reflabeL#2%
\ifWritingAuxFile\immediate\write\auxfile{\noexpand\xdef\noexpand#2{#2}}\fi%
}

\def\fign#1{\makeNormalizedRomappno%
\ifnum\secno>0%
  \warnIfChanged#1{\the\secno.\char\the\mfino}%
  {\ninebf Figure \the\secno.\char\the\mfino:}\xdef#1{\the\secno.\char\the\mfino}%
     \global\advance\mfino by1
\else\ifnum\appno>0%
  \warnIfChanged#1{\normalizedromappno.\char\the\mfino}%
  {\ninebf Figure \romappno.\char\the\mfino}%
      \xdef#1{\normalizedromappno.\char\the\mfino}%
     \global\advance\mfino by1
\else%
  \warnIfChanged#1{\char\the\mfino}%
 {\ninebf Figure \char\the\mfino:}\xdef#1{\char\the\mfino}%
     \global\advance\mfino by1
  \fi\fi%
\ifWritingAuxFile\immediate\write\auxfile{\noexpand\xdef\noexpand#1{#1}}\fi%
}

\catcode`@=11 

\font\ninerm=cmr9
\font\eightrm=cmr8
\font\sixrm=cmr6

\def\loadtrueseventeenpoint{
 \font\seventeenrm=cmr10 at 17.28truept
 \font\seventeeni=cmmi10 at 17.28truept
 \font\seventeenbf=cmbx10 at 17.28truept
 \font\seventeenit=cmti10 at 17.28truept
 \font\seventeensl=cmsl10 at 17.28truept
 \font\seventeensy=cmsy10 at 17.28truept
}
\def\loadfourteenpoint{
\font\fourteenrm=cmr10 at 14.4pt
\font\fourteeni=cmmi10 at 14.4pt
\font\fourteenit=cmti10 at 14.4pt
\font\fourteensl=cmsl10 at 14.4pt
\font\fourteensy=cmsy10 at 14.4pt
\font\fourteenbf=cmbx10 at 14.4pt
}
\def\loadtruetwelvepoint{
\font\twelverm=cmr10 at 12truept
\font\twelvei=cmmi10 at 12truept
\font\twelveit=cmti10 at 12truept
\font\twelvesl=cmsl10 at 12truept
\font\twelvesy=cmsy10 at 12truept
\font\twelvebf=cmbx10 at 12truept
}

\font\ninei=cmmi9
\font\eighti=cmmi8
\font\sixi=cmmi6
\skewchar\ninei='177 \skewchar\eighti='177 \skewchar\sixi='177

\font\ninesy=cmsy9
\font\eightsy=cmsy8
\font\sixsy=cmsy6
\skewchar\ninesy='60 \skewchar\eightsy='60 \skewchar\sixsy='60

\font\ninebf=cmbx9
\font\eightbf=cmbx8
\font\sixbf=cmbx6

\font\ninett=cmtt9
\font\eighttt=cmtt8

\hyphenchar\tentt=-1 
\hyphenchar\ninett=-1
\hyphenchar\eighttt=-1

\font\ninesl=cmsl9
\font\eightsl=cmsl8

\font\nineit=cmti9
\font\eightit=cmti8


\newskip\ttglue
\def\tenpoint{\def\rm{\fam0\tenrm}%
  \textfont0=\tenrm \scriptfont0=\sevenrm \scriptscriptfont0=\fiverm
  \textfont1=\teni \scriptfont1=\seveni \scriptscriptfont1=\fivei
  \textfont2=\tensy \scriptfont2=\sevensy \scriptscriptfont2=\fivesy
  \textfont3=\tenex \scriptfont3=\tenex \scriptscriptfont3=\tenex
  \def\it{\fam\itfam\tenit}\textfont\itfam=\tenit
  \def\sl{\fam\slfam\tensl}\textfont\slfam=\tensl
  \def\bf{\fam\bffam\tenbf}\textfont\bffam=\tenbf \scriptfont\bffam=\sevenbf
  \scriptscriptfont\bffam=\fivebf
  \normalbaselineskip=12pt
  \let\sc=\eightrm
  \let\big=\tenbig
  \setbox\strutbox=\hbox{\vrule height8.5pt depth3.5pt width\z@}%
  \normalbaselines\rm}

\def\twelvepoint{\def\rm{\fam0\twelverm}%
  \textfont0=\twelverm \scriptfont0=\ninerm \scriptscriptfont0=\sevenrm
  \textfont1=\twelvei \scriptfont1=\ninei \scriptscriptfont1=\seveni
  \textfont2=\twelvesy \scriptfont2=\ninesy \scriptscriptfont2=\sevensy
  \textfont3=\tenex \scriptfont3=\tenex \scriptscriptfont3=\tenex
  \def\it{\fam\itfam\twelveit}\textfont\itfam=\twelveit
  \def\sl{\fam\slfam\twelvesl}\textfont\slfam=\twelvesl
  \def\bf{\fam\bffam\twelvebf}\textfont\bffam=\twelvebf
                                            \scriptfont\bffam=\ninebf
  \scriptscriptfont\bffam=\sevenbf
  \normalbaselineskip=12pt
  \let\sc=\eightrm
  \let\big=\tenbig
  \setbox\strutbox=\hbox{\vrule height8.5pt depth3.5pt width\z@}%
  \normalbaselines\rm}

\def\fourteenpoint{\def\rm{\fam0\fourteenrm}%
  \textfont0=\fourteenrm \scriptfont0=\tenrm \scriptscriptfont0=\sevenrm
  \textfont1=\fourteeni \scriptfont1=\teni \scriptscriptfont1=\seveni
  \textfont2=\fourteensy \scriptfont2=\tensy \scriptscriptfont2=\sevensy
  \textfont3=\tenex \scriptfont3=\tenex \scriptscriptfont3=\tenex
  \def\it{\fam\itfam\fourteenit}\textfont\itfam=\fourteenit
  \def\sl{\fam\slfam\fourteensl}\textfont\slfam=\fourteensl
  \def\bf{\fam\bffam\fourteenbf}\textfont\bffam=\fourteenbf%
  \scriptfont\bffam=\tenbf
  \scriptscriptfont\bffam=\sevenbf
  \normalbaselineskip=17pt
  \let\sc=\elevenrm
  \let\big=\tenbig
  \setbox\strutbox=\hbox{\vrule height8.5pt depth3.5pt width\z@}%
  \normalbaselines\rm}

\def\seventeenpoint{\def\rm{\fam0\seventeenrm}%
  \textfont0=\seventeenrm \scriptfont0=\fourteenrm \scriptscriptfont0=\tenrm
  \textfont1=\seventeeni \scriptfont1=\fourteeni \scriptscriptfont1=\teni
  \textfont2=\seventeensy \scriptfont2=\fourteensy \scriptscriptfont2=\tensy
  \textfont3=\tenex \scriptfont3=\tenex \scriptscriptfont3=\tenex
  \def\it{\fam\itfam\seventeenit}\textfont\itfam=\seventeenit
  \def\sl{\fam\slfam\seventeensl}\textfont\slfam=\seventeensl
  \def\bf{\fam\bffam\seventeenbf}\textfont\bffam=\seventeenbf%
  \scriptfont\bffam=\fourteenbf
  \scriptscriptfont\bffam=\twelvebf
  \normalbaselineskip=21pt
  \let\sc=\fourteenrm
  \let\big=\tenbig
  \setbox\strutbox=\hbox{\vrule height 12pt depth 6pt width\z@}%
  \normalbaselines\rm}

\def\ninepoint{\def\rm{\fam0\ninerm}%
  \textfont0=\ninerm \scriptfont0=\sixrm \scriptscriptfont0=\fiverm
  \textfont1=\ninei \scriptfont1=\sixi \scriptscriptfont1=\fivei
  \textfont2=\ninesy \scriptfont2=\sixsy \scriptscriptfont2=\fivesy
  \textfont3=\tenex \scriptfont3=\tenex \scriptscriptfont3=\tenex
  \def\it{\fam\itfam\nineit}\textfont\itfam=\nineit
  \def\sl{\fam\slfam\ninesl}\textfont\slfam=\ninesl
  \def\bf{\fam\bffam\ninebf}\textfont\bffam=\ninebf \scriptfont\bffam=\sixbf
  \scriptscriptfont\bffam=\fivebf
  \normalbaselineskip=11pt
  \let\sc=\sevenrm
  \let\big=\ninebig
  \setbox\strutbox=\hbox{\vrule height8pt depth3pt width\z@}%
  \normalbaselines\rm}

\def\eightpoint{\def\rm{\fam0\eightrm}%
  \textfont0=\eightrm \scriptfont0=\sixrm \scriptscriptfont0=\fiverm%
  \textfont1=\eighti \scriptfont1=\sixi \scriptscriptfont1=\fivei%
  \textfont2=\eightsy \scriptfont2=\sixsy \scriptscriptfont2=\fivesy%
  \textfont3=\tenex \scriptfont3=\tenex \scriptscriptfont3=\tenex%
  \def\it{\fam\itfam\eightit}\textfont\itfam=\eightit%
  \def\sl{\fam\slfam\eightsl}\textfont\slfam=\eightsl%
  \def\bf{\fam\bffam\eightbf}\textfont\bffam=\eightbf \scriptfont\bffam=\sixbf%
  \scriptscriptfont\bffam=\fivebf%
  \normalbaselineskip=9pt%
  \let\sc=\sixrm%
  \let\big=\eightbig%
  \setbox\strutbox=\hbox{\vrule height7pt depth2pt width\z@}%
  \normalbaselines\rm}

\def\tenbig#1{{\hbox{$\left#1\vbox to8.5pt{}\right.\n@space$}}}
\def\ninebig#1{{\hbox{$\textfont0=\tenrm\textfont2=\tensy
  \left#1\vbox to7.25pt{}\right.\n@space$}}}
\def\eightbig#1{{\hbox{$\textfont0=\ninerm\textfont2=\ninesy
  \left#1\vbox to6.5pt{}\right.\n@space$}}}

\def\footnote#1{\edef\@sf{\spacefactor\the\spacefactor}#1\@sf
      \insert\footins\bgroup\eightpoint
      \interlinepenalty100 \let\par=\endgraf
        \leftskip=\z@skip \rightskip=\z@skip
        \splittopskip=10pt plus 1pt minus 1pt \floatingpenalty=20000
        \smallskip\item{#1}\bgroup\strut\aftergroup\@foot\let\next}
\skip\footins=12pt plus 2pt minus 4pt 
\dimen\footins=30pc 

\newinsert\margin
\dimen\margin=\maxdimen
\def\titlefont{\seventeenpoint}
\loadtruetwelvepoint 
\loadtrueseventeenpoint

\def\eatOne#1{}
\def\ifundef#1{\expandafter\ifx%
\csname\expandafter\eatOne\string#1\endcsname\relax}
\def\notTrue{\iffalse}\def\isTrue{\iftrue}
\def\ifdef#1{{\ifundef#1%
\aftergroup\notTrue\else\aftergroup\isTrue\fi}}
\def\use#1{\ifundef#1\linemessage{Warning: \string#1 is undefined.}%
{\tt \string#1}\else#1\fi}


\global\newcount\refno \global\refno=1
\newwrite\rfile
\newlinechar=`\^^J
\def\@ref#1#2{\the\refno\n@ref#1{#2}}
\def\n@ref#1#2{\xdef#1{\the\refno}%
\ifnum\refno=1\immediate\openout\rfile=\jobname.refs\fi%
\immediate\write\rfile{\noexpand\item{[\noexpand#1]\ }#2.}%
\global\advance\refno by1}
\def\nref{\n@ref} 
\def\ref{\@ref}   
\def\lref#1#2{\the\refno\xdef#1{\the\refno}%
\ifnum\refno=1\immediate\openout\rfile=\jobname.refs\fi%
\immediate\write\rfile{\noexpand\item{[\noexpand#1]\ }#2\semi}%
\global\advance\refno by1}
\def\cref#1{\immediate\write\rfile{#1\semi}}

\def\preref#1#2{\gdef#1{\@ref#1{#2}}}

\def\semi{;\hfil\noexpand\break}

\def\listrefs{\vfill\eject\immediate\closeout\rfile
\centerline{{\twelvepoint\bf References}}\medskip\frenchspacing%
\input \jobname.refs\vfill\eject\nonfrenchspacing}

\def\inputAuxIfPresent#1{\immediate\openin1=#1
\ifeof1\message{No file \auxfileName; I'll create one.
}\else\closein1\relax\input\auxfileName\fi%
}

\newif\ifWritingAuxFile
\newwrite\auxfile
\def\SetUpAuxFile{%
\xdef\auxfileName{\jobname.aux}%
\inputAuxIfPresent{\auxfileName}%
\WritingAuxFiletrue%
\immediate\openout\auxfile=\auxfileName}



\catcode`\@=\active
\catcode`@=12  
\catcode`\"=\active


\def\h{{\hbox{$1\over2$}}}

\def\spa#1.#2{\left\langle#1\,#2\right\rangle}
\def\spb#1.#2{\left[#1\,#2\right]}
\def\lor#1.#2{\left(#1\,#2\right)}
\def\sand#1.#2.#3{%
\left\langle\smash{#1}{\vphantom1}^{-}\right|{#2}%
\left|\smash{#3}{\vphantom1}^{-}\right\rangle}
\catcode`@=11  
\def\meqalign#1{\,\vcenter{\openup1\jot\m@th
   \ialign{\strut\hfil$\displaystyle{##}$ && $\displaystyle{{}##}$\hfil
             \crcr#1\crcr}}\,}
\catcode`@=12  


\SetUpAuxFile
\loadfourteenpoint
\hfuzz 60 pt


\def\ref{\preref}
\def\cut{\bigskip\noindent}

\def\eps{\epsilon}
\def\Atree{A^{\rm tree}}
\def\overlrarrow#1{\vbox{\ialign{##\crcr$\leftrightarrow$\crcr
\noalign{\kern0pt
\nointerlineskip}
$\hfil\displaystyle{#1}\hfil$\crcr}}}


\input epsf.tex


\ref\GravityReview{M.  Veltman, in {\it Les Houches 1975, Methods in Field
Theory}, ed R. Balian and J. Zinn-Justin, (North Holland, Amsterdam,
1976)\semi
R.P. Feynman, {\it Lectures on Gravitation} (Addison-Wesley, 1995)}
\ref\SUGRARev{P. van Nieuwenhuizen, Phys.\ Rep.\ 68:189 (1981)}

\ref\HVb{G. 't\ Hooft and M.\ Veltman, Ann. Inst. Henri Poincar\'e 20:69
(1974)}

\ref\Background{G. 't Hooft, Acta Universitatis Wratislavensis no.\ 38, 12th
Winter School of Theoretical Physics in Karpacz; {\it Functional and
Probabilistic Methods in Quantum Field Theory}, Vol. 1 (1975)\semi 
 B.S.\ DeWitt, in {\it Quantum Gravity II}, eds. C. Isham, R.\ Penrose 
and D.\ 
Sciama (Oxford, 1981)\semi 
 L.F.\ Abbott, Nucl.\ Phys.\ B185:189 (1981)\semi 
 L.F\ Abbott, M.T.\ Grisaru and R.K.\ Schaeffer, Nucl.\ Phys. {B229}:372 
(1983)\semi
{Z. Bern and D.C.\ Dunbar,  Nucl.\ Phys.\ B379:562 (1992)}
 }
\ref\EMGrav{S. Deser and  P. van Nieuwenhuizen,
Phys.\ Rev.\ D10:401 (1974)\semi
M.T. Grisaru, P. van Nieuwenhuizen and  C.C.\ Wu,
Phys.\ Rev.\ D12:1813 (1975)}   

\ref\YMGrav{S. Deser,  Hung-Sheng Tsao and  P. van Nieuwenhuizen,
Phys.\ Lett.\ 50B:491 (1974); 
Phys.\ Rev.\ D10:3337 (1974)}
\ref\FermCt{S. Deser and  P. van Nieuwenhuizen,
Phys.\ Rev.\ D10:411 (1974)}

\ref\Cutkosky{L.D.\ Landau, Nucl.\ Phys.\ 13:181 (1959)\semi
 S. Mandelstam, Phys.\ Rev.\ 112:1344 (1958), 115:1741 (1959)\semi
 R.E.\ Cutkosky, J.\ Math.\ Phys.\ 1:429 (1960)}

\ref\CutRefs{Z. Bern, D.C. Dunbar, L. Dixon and D.A. Kosower, 
Nucl.\ Phys.\ B435:59 (1995)\semi
Z. Bern, D.C. Dunbar, L. Dixon and D.A. 
Kosower, Nucl.\ Phys.\ B425:217 (1994)}

\ref\ScalCut{D.C.\  Dunbar and P.S.\ Norridge, hep-th/9512084}

\ref\StringBased{ Z. Bern and D.A.\ Kosower,
Nucl.\ Phys.\ B379:451 (1992)}

\ref\SusyA{ M.T. Grisaru, H.N. Pendleton and P.  van Nieuwenhuizen, 
Phys. Rev. {D15}:996 (1977)}
\ref\SusyB{ M.T. Grisaru and  H.N. Pendleton, Nucl. Phys. B124:81 (1977)\semi
M.L.\ Mangano and S.J. Parke, Phys.\ Rep.\ {200}:301 (1991)}
\ref\SusyC{S.J. Parke and T.R. Taylor, Phys.Lett.157B:81 (1985) 
[Erratum: 174B:465 (1986)]}

\ref\XZC{
and T.T
.\ Wu, Phys.\ Lett.\ 103B:124 (1981)\semi
 P.\ De Causmaeker, R.\ Gastmans, W.\ Troost and T.T.\ Wu, Nucl. Phys. 
B206:53
(1982)\semi
 R.\ Kleiss and W.\ J.\ Stirling, Nucl.\ Phys.\ B262:235 (1985)\semi
 J.\ F.\ Gunion and Z.\ Kunszt, Phys.\ Lett.\ 161B:333 (1985)\semi
 R.\ Gastmans and T.T.\ Wu, {\it The Ubiquitous Photon: Helicity Method 
for QED
and QCD} (Clarendon Press) (1990)\semi
 Z.\ Xu, D.-H.\ Zhang and L. Chang, Nucl.\ Phys.\ B291:392 (1987)}

\ref\Berends{F.A. Berends, W.T.\ Giele and H. Kuijf, Phys. Lett.\ {211B}:91
(1988)}

\ref\Sannan{S.\ Sannan, Phys.\ Rev.\ D34:1748 (1986)}

\ref\TwoQuark{Z. Bern, L. Dixon and D.A. Kosower,  Nucl.\ Phys.\ B437:259 
(1995)
}

\ref\GravityString{D.C.\  Dunbar and P.S.\ Norridge,
Nucl.\ Phys.\ B433:181 (1995)}

\nopagenumbers
\loadfourteenpoint 

\noindent

$\null$

\vskip -1.6 cm

hep-th/9606067

\rightline{SWAT-96-100}

\hfill June 1996


\vskip 1in
{\titlefont\centerline{\bf Recovering Infinities in Graviton Scattering }}
{\titlefont\centerline{\bf   Amplitudes using Cutkosky rules.}} \vskip .3cm
\vskip .5 in

\centerline{\bf Paul S. Norridge\footnote{$^\dagger$}{email: p.s.norridge@swan.ac.uk}} 
\centerline{\it Department of Physics}
\centerline{\it University of Wales, Swansea}
\centerline{\it Swansea SA2 8PP}
\centerline{\it UK }
        
\vskip 1.2 truecm \baselineskip12pt

\centerline{\bf Abstract } { \narrower\smallskip \smallskip 
We use the Cutkosky rules as a tool for determining the
infinities present in graviton scattering amplitudes.
We are able to confirm theoretical derivations of counterterms in
Einstein-Maxwell theory and to determine new results in the 
Dirac-Einstein counter-Lagrangian.
  }

\baselineskip14pt

\vfill\eject

\footline={\hss\tenrm\folio\hss}

\noindent

One of the central issues in perturbative quantum
gravity~[\use\GravityReview] and supergravity~[\use\SUGRARev] is that of
renormalisation. The dimensionful coupling constant in such theories
implies that new infinities could be introduced at each perturbative
order. Each new infinity would require new terms to be added to the
counter-Lagrangian.  Clearly, the explicit evaluation of ultra-violet
infinities is a necessary step in determining the structure of these
theories. 

In some cases simple arguments about the symmetries of a theory can show
that all counterterms vanish (for example, for on-shell pure gravity at
one loop). However, in general, the counter-Lagrangian can only be found
by calculation.  Unfortunately, the well-known difficulties involved in
perturbative gravity calculations impede such investigations
significantly.  Calculations {\it have} been done at one-loop for gravity
coupled to matter and at two-loops for pure gravity.  Where that matter is
purely bosonic the full counter-Lagrangian can be determined using an
algorithm due to `t Hooft and Veltman~[\use\HVb].  This method examines
counterterms in the effective action using the background field
method~[\use\Background]. It has been used successfully to examine gravity
coupled to scalars~[\use\HVb] and spin-one
particles~[\use\EMGrav,\use\YMGrav]. In all cases the theories were shown
to be non-renormalisable.  The `t Hooft-Veltman algorithm is not so useful
in theories involving fermions.  However, in the Dirac-Einstein system
some progress has been made via direct calculation~[\use\FermCt]; by
focussing on the infinities in a well-chosen amplitude it was possible to
determine one of the counterterms.  While this is sufficient to show that
the theory is non-renormalisable, no expression for the complete
counter-Lagrangian has been constructed.  In supergravity theories
three-loop results are required to decide renormalisability; due to the
difficulty of the calculations involved, such results remain undetermined. 
In this paper we will determine ultra-violet infinities directly from 
physical amplitudes using the Cutkosky rules.

The Cutkosky rules~[\use\Cutkosky,\use\CutRefs] are a useful way to obtain
information about amplitudes; they allow us to construct expressions
containing the correct cuts in all channels.  In a recent
paper~[\use\ScalCut] it was shown how these rules can provide information
about the one-loop divergences:  Since logarithmic terms can be found
exactly from the Cutkosky-based calculations, and these only appear as
part of the expansion
 $$
 (-s_{ij})^{-\eps} = 1 - \epsilon\ln(-s_{ij}) 
+ \h{\epsilon^{2}}\ln(-s_{ij})^2+ \cdots 
\anoneqn
$$
 at one-loop, it is a straightforward task to deduce all infinite
contributions. (We work in the `four dimensional helicity' form of
dimensional regularisation~[\use\StringBased] with $\eps = 2-D/2$.) Since
the IR divergences can be derived independently, this method can be used
to identify one-loop UV divergences in a relatively efficient way.  This
was used to confirm previous derivations of counterterms for gravity
coupled to scalars.  In this paper we will apply the same method to cases
with gravity coupled to fermions and photons.  In the former case this
will enable us to find new information about the Dirac-Einstein
counter-Lagrangian.  In the latter we will show how the supersymmetric
Ward identities~[\use\SusyA,\use\SusyB,\use\SusyC] can be used to further 
simplify
this method; our result will confirm the previous derivation of the
Einstein-Maxwell counterterms.


\goodbreak

\

\centerline{\it Gravity Coupled to Fermions}
\medskip
\noindent
 Let us begin with the theory with gravity coupled to
fermions~[\use\FermCt].  The process for obtaining cuts in this case
follows the scalar calculation~[\use\ScalCut] very closely. The simplest
case we can consider is the amplitude with four external fermions.  Rather
than calculate this for a general case, let us look at specific choices of
external particle helicity. There are three independent helicity
configurations which can be considered:  $(+,+,+,+)$, $(-,+,+,+)$ and
$(-,+,-,+)$.  It is easy to show that all cuts vanish for the first two,
implying that these amplitudes will be divergence free, so let us
concentrate on the case $(1^-,2^+,3^-,4^+)$ (with chiral fermions).  

For a full cut calculation we must sum over all internal states.  In this
cases that means including both graviton and fermion contributions
(fig.~\use\FourScal) and considering all internal helicity configurations.
Since there is an asymmetry amongst the external helicities which must
also look for different contributions in different channels. (In fact we 
need only consider the  the $u$- and $s$-channels; the `$s$' and
`$t$' calculations {\it will} be related by symmetry.)

\midinsert
{\smallskip 
\centerline{\epsfxsize=2.5in \epsfbox{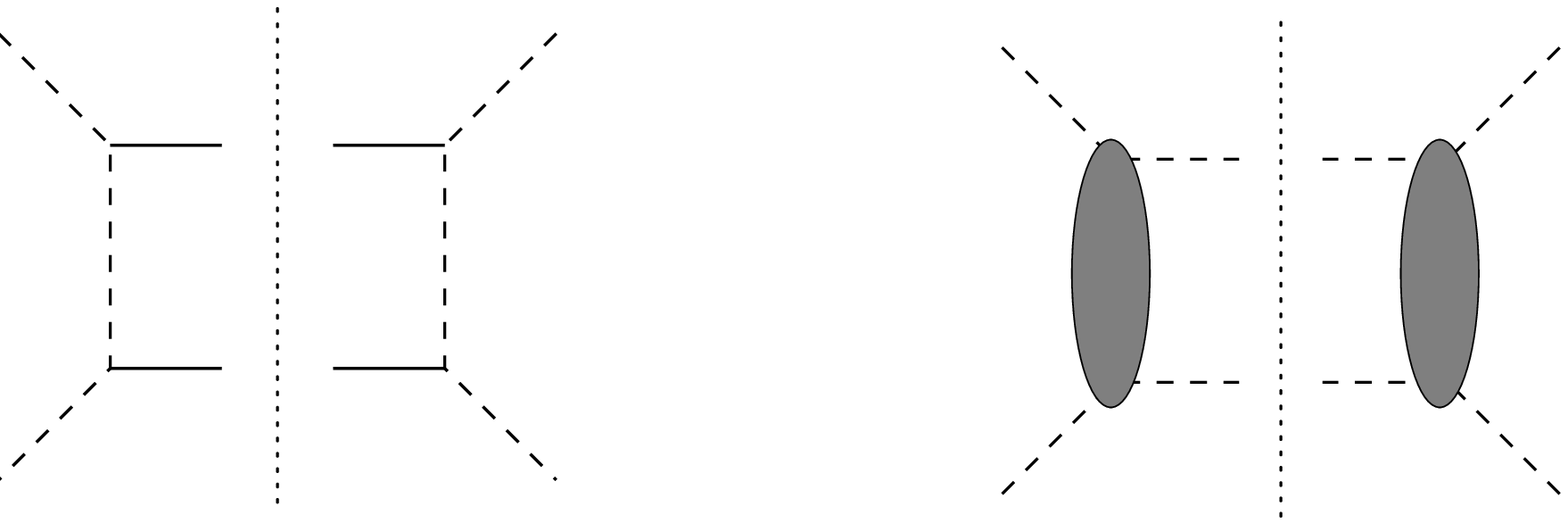}}
\vskip -0.1in 
\centerline{\eightrm(a)\hskip 1.45in (b)}
\smallskip 
{\baselineskip 10 pt\ninerm
\centerline{{\fign\FourScal} Cut contributions required for the}
\centerline{four fermion calculation. } 
}}
\endinsert 

 First, consider the $u$-channel. In this case the internal graviton
contribution is trivial; there is no internal helicity configuration for
which all the trees involved in the calculation are non-zero. So, we only
need to calculate cuts with internal fermions. The building block for this
case will be the tree with four external fermions.  The only choice of
external helicities for which the tree is non-zero is $\Atree(-,+,-,+)$.  We
can find this from direct calculation: 
 $$\eqalign{
 A(1^-,2^+,3^-,4^+) &= -{i\spa1.3\spb2.4\kappa^2\over 8\,st}(2s^2+st+2t^2).
} 
\eqn\FermTree
$$
 (Note that throughout this work we use the spinor-helicity notation of 
refs.~[\use\XZC,\use\SusyB].)  Sewing two such trees together, expanding 
this and integrating gives us the $u$-channel cut contribution to be: 
 $$\eqalign{-{ir_\Gamma\,\kappa^4\over(4\pi)^{2-\eps}}
   {1\over16\spa2.4\spb1.3}\bigg(
&{ {u^{3}\left (2t^{2}+st+2s^{2}\right )\ln (u)}\over{st}}{1\over\eps}
+{ {2\,u^{4}\ln (u)\ln (t)}\over{t}}
\cr&\qquad\qquad
+{{2\,u^{4}\ln (u)\ln (s)}\over{s}}
+{3\over2}u^{3}\ln (u)^2
-{9\over8}\,u^{3}\ln (u)
 \bigg).
}
\anoneqn
$$

The $s$-channel has contributions from both particle types in the loop. 
The calculation for the fermion contribution is very similar to the
$u$-channel case; as before (\use\FermTree) is the basic building block.
Using this we find the cut contribution to be
 $$\eqalign{
 -{i\kappa^4\,{r_\Gamma}\over(4\pi)^{2-\eps}}&{1\over16\spa2.4\spb1.3}
\bigg({1\over\eps}{{u^{2}\left(2u^{2}+3ut+3t^{2}\right )}\over{t}}\ln(s)
-{{s^{2}\left (2\,u^{2}-t^{2}\right)}\over t}\ln(s)\ln (t)
\cr&\qquad
-{{\left (t^{3}+2ut^{2}+2u^{2}t-u^{3}\right )}\over2}\ln (s)^{2}
+{{u\left(274\,ut-53\,u^{2}+60\,t^{2}\right)}\over {60}}\ln (s)
\bigg).\cr
}
\eqn\FermFermRes
$$

 The tree used for the cut with internal gravitons can be found using a
previous result for the pure gravity tree~[\use\Berends,\use\Sannan] and
SUSY Ward Identities.  We can write it as
 $$\eqalign{ 
A(g^-,f^-,f^+,g^+) &=
{{\spa1.3}^3\over{\spa1.2}^3} A(g^-,g^-,g^+,g^+)
= {i\kappa^2\over4} {{\spa1.2}^3{\spa1.3}^3\over
                       {\spa2.3}^2{\spa3.4}^2{\spa1.4}^2}
{st\over u}.
}
\anoneqn
$$
Again, sewing two trees of this type together we obtain a cut result of
 $$\eqalign{
 -{i\kappa^4{r_\Gamma}\over(4\pi)^{2-\eps}}
&{1\over\spa2.4\spb1.3} \bigg({{u^{4}\ln (s)\ln (u)}\over{8\,s}}
-{1\over32}\left (4\,t^{3}+2\,s^{3}+11\,t^{2}s+8\,ts^{2}\right )
\ln (s)^{2}
\cr
&-{{t^{2}\left
(2\,ts+2\,t^{2}+s^{2}\right )\ln (s)\ln (t)}\over{16\,s}}-{1\over960}
{{\,u\left (7\,s^{2}-5\,ts+48\,t^{2}\right )\ln (s)}}\bigg).
}
\eqn\FermGravRes
$$
Summing (\use\FermFermRes) and (\use\FermGravRes), we see that the
total $s$-channel
cuts are
$$\eqalign{-{i\kappa^4{r_\Gamma}\over(4\pi)^{2-\eps}}&{u\over16\spa2.4\spb1.3}
\bigg({1\over\eps}{{u^{2}\left (2u^{2}+3ut+3t^{2}\right )}\over{t}}\ln(s)
+{1\over4}{{u\left (21s+4\,t\right )\ln (s)}}
\cr&\quad
+{2\,{u^{3}\ln (s)\ln (u)}\over{s}}
+{{
\left (2\,s^{4}+2\,ts^{3}-t^{2}s^{2}+2\,st^{3}+2\,t^{4}\right )\ln (s)
\ln (t)}\over{st}}
\cr&\quad\qquad\qquad\qquad\qquad\qquad\qquad\qquad\qquad
+{1\over2}\left (3\,s^{2}+10\,st+6\,t^{2}\right
)\ln (s)^{2}
\bigg).
}
\anoneqn
$$
 We can find the $t$-channel contribution from this by making the exchange
$s\leftrightarrow t$. If we combine the results from the three channels we
can deduce the total expression. As expected, we have an IR contribution
 $$
-{i\kappa^4{r_\Gamma}\over2\,(4\pi)^{2-\eps}}{1\over\eps}
\left(u\ln(u)+s\ln(s)+t\ln(t)\right){1\over8}
{ {u^2\left (st+2t^{2}+2s^{2}\right
)}\over{\spa2.4\spb1.3}{st}}
 \anoneqn
$$
(see ref.~[\use\ScalCut] for details of the identification of IR 
contributions). 
Removing this leaves the UV component easily identifiable as
$$
-{59\over128}{i\kappa^4{r_\Gamma}\over(4\pi)^{2}}{1\over\eps}
{u^3\over\spa2.4\spb1.3}.
\eqn\FermRes
$$
\cut
This expression must be cancelled by a four-point counterterm.  This case
has not been considered before, so we can use our result to deduce new
information about the Dirac-Einstein counter-Lagrangian.  The general
on-shell counterterm which we must consider takes the form~[\use\FermCt]
 $$
{\alpha\over2}\,
{\kappa^4\over(4\pi)^{2}}
{e\over\eps}
\left(\bar\eta\gamma_\mu\overlrarrow\partial_\nu\eta\right)^2.
\anoneqn
$$
 where $e$ is the determinant of the vierbein and $\alpha$ is a constant
to be determined. (Note that other all other possibilities can be related
to this by a combination of on-shell conditions, integration by parts and
the Fierz theorem.)
We can deduce that this will give a counterterm contribution of
$$\eqalign{ {i\kappa^4\over(4\pi)^{2}}{\alpha\over\eps}
&\Big(\left\langle2^+|\gamma_\mu|1^+\right\rangle
     \left\langle4^+|\gamma^\mu|3^+\right\rangle\,(u-t)
   -\left\langle4^+|\gamma_\mu|1^+\right\rangle
     \left\langle2^+|\gamma^\mu|3^+\right\rangle\,(u-s)\Big)\cr
&\quad
\qquad\qquad\qquad=-6\,\alpha
{i\kappa^4\over(4\pi)^{2}}{u^3\over\spa2.4\spb1.3}{1\over\eps}. }
\anoneqn
$$
Comparing this with (\use\FermRes) implies that
$$
\alpha = -{59\over768}.
\anoneqn
$$
 The fact that this coefficient is non-zero is confirmation
that the Dirac-Einstein system is non-renormalisable.  Note that we have
evaluated part of the counter-Lagrangian which could not be determined
from the calculation in ref.~[\FermCt].

\


\goodbreak
\centerline{\it Gravity Coupled to Photons}
 
\medskip
\noindent
 We could take the same route as above to find the infinities in photon
amplitudes, but we can reduce the work by a judicious use of SUSY
identities. (These have also proved useful in gauge theories one-loop 
computations~[\use\TwoQuark].) 
Let us look here at the four point amplitude with no external
gravitons.  Although our aim is to find the infinities in a system
containing photons and gravitons, let us begin by considering the $N=2$
supersymmetric multiplet containing a photon a graviton and two
gravitinos.  The cut for this amplitude is the sum of the cuts with
photons, gravitinos and gravitons in the loop, fig.~\use\PhotA.

\midinsert
{\medskip
\centerline{\epsfxsize=3in \epsfbox{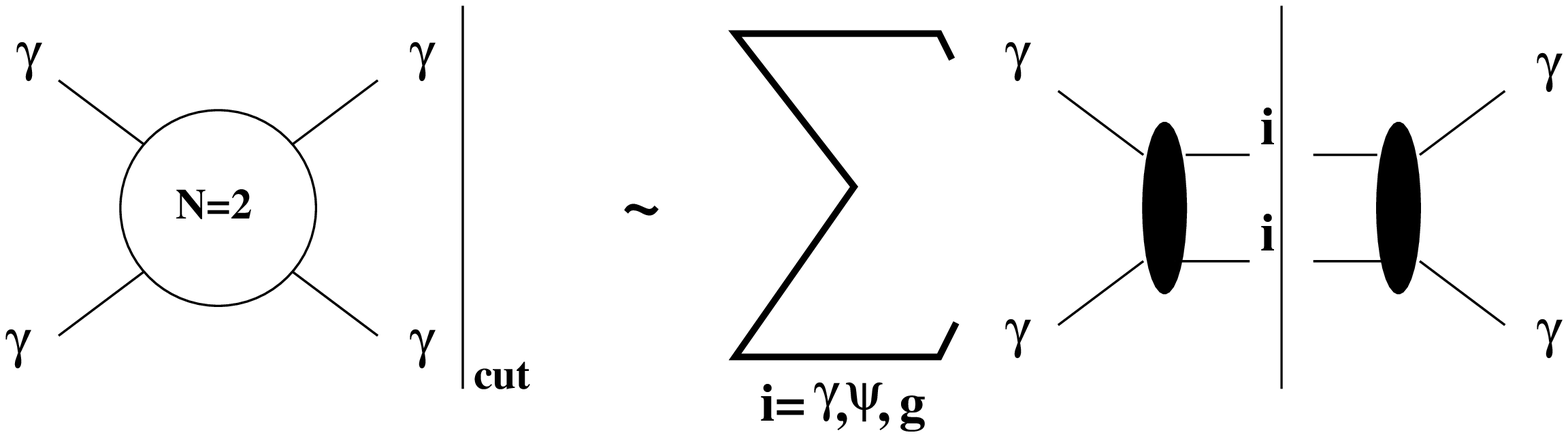}}
\smallskip 
{\baselineskip 10 pt\ninerm
\centerline{\fign\PhotA\ Cut contributions required for the one loop}
\centerline{ four photon amplitude in an $N=2$ SUGRA theory.}
}}
\endinsert

\goodbreak

Notice that the tree amplitudes $\Atree(\gamma,\gamma,\gamma,\gamma)$ and
$\Atree(g , g , g, g)$ do not include any contribution due to gravitinos
(since any internal gravitinos must form a complete loop). Hence, the only
gravitino contribution is contained in the second term in the sum in
fig.~\use\PhotA. We can deduce that the cuts for the amplitude in which we
are interested, the one involving only gravitons and photons, are equal to
the cuts of the $N=2$ amplitude minus the cut containing 
gravitinos, fig.~\use\PhotB. 

\smallskip

\midinsert
{\smallskip
\centerline{\epsfxsize=4in \epsfbox{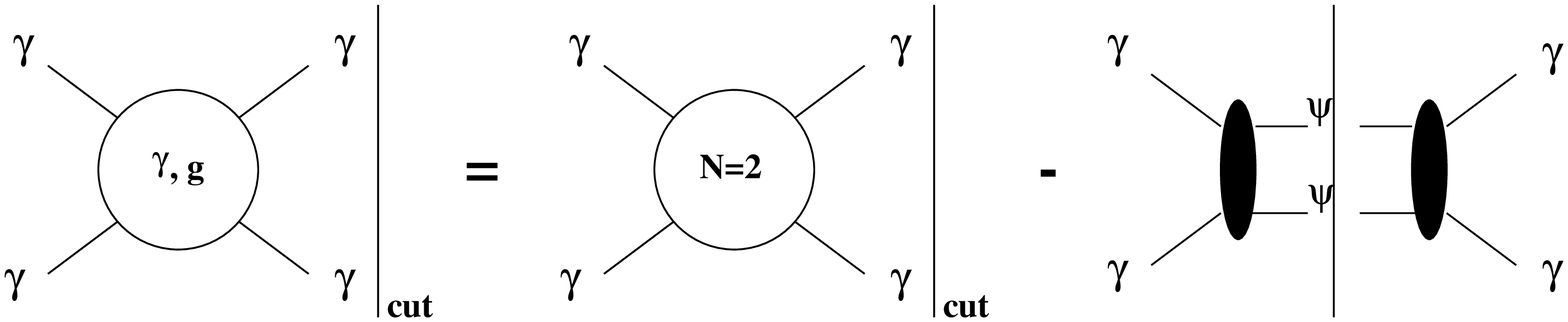}}
\smallskip
{\baselineskip 10 pt\ninerm
\centerline{\fign\PhotB\ Cut equation for the one loop four photon }
\centerline{amplitude in Einstein-Maxwell theory.}
}}
\endinsert

Now, using supersymmetric relations between amplitudes we can make the
following deductions: First, for choices of external helicities
$(\gamma^{-} \gamma^{+} \gamma^{+} \gamma^{+})$ and $(\gamma^{+}
\gamma^{+} \gamma^{+} \gamma^{+})$ the gravitino contributions vanish and
the cuts will be precisely equal to the cuts in the $N$ amplitude.  We
know the $N=2$ amplitude contains only IR divergences and hence there
will be no UV infinite components in the photon-graviton amplitudes with
those helicity choices. 

Since we are only looking for UV divergences, the helicity configuration of
interest to us here is $A(\gamma^{-} \gamma^{-} \gamma^{+} \gamma^{+})$.
SUSY relations give us
 $$
A^{\rm N=2}(\gamma^{-} \gamma^{-} \gamma^{+}
\gamma^{+}) = {\spa4.3^2 \over \spa1.2^2}A^{\rm N=2}(g^{-} g^{-} g^{+}
g^{+}),
\anoneqn
$$
 implying the relation in fig.~\use\PhotC. 

\midinsert
{\centerline{\epsfxsize=4in \epsfbox{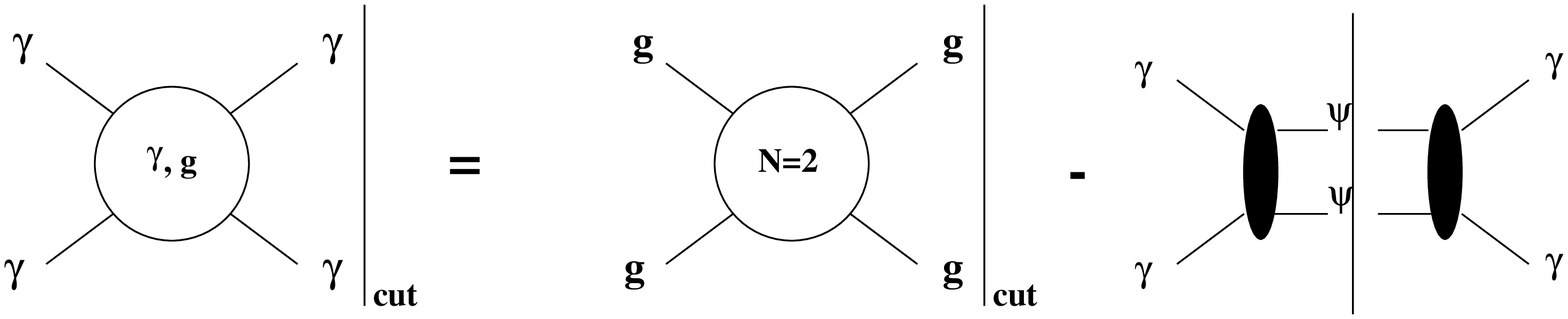}}
\vskip -0.6in
\centerline{\hbox{\vbox{\fiverm$${\spa4.3^2 \over \spa1.2^2}\times$$}
\hskip.9in}} \vskip 0.1in
{\baselineskip 10 pt\ninerm
\centerline{\fign\PhotC\ Cut equation for the one loop four photon }
\centerline{amplitude in Einstein-Maxwell theory (rewritten).}
}}
\endinsert

The one-loop amplitude $A^{\rm N=2}(g^{-} g^{-} g^{+} g^{+})$ has been
calculated in ref.~[\use\GravityString].  Using this we find that the cut
contribution from the first term on the right hand side is
 $$
\eqalign{
 F' \biggl( &
 {2 \over \epsilon } \bigg(  {\ln(-u) \over st} + {\ln(-t) \over su}
+  {\ln(-s) \over tu}  \biggr)
\cr &\qquad+
{{2\,\ln (-u)\ln (-s)}\over{s
u}}
+{ {2\,\ln (-t)\ln (-u)}\over{tu}}
+{{2\,\ln (-t)\ln (-s)}\over{ts}}\cr
&\qquad\qquad+{ {\left (   
3\,t^{4}+3\,t
^{3}u-t^{2}u^{2}+3\,tu^{3}+3\,u^{4}
\right )  
(\ln^2 ({-t/-u})) }\over{s^{6}}}\cr
&\qquad\qquad\qquad\qquad+{ {\left (t-u\right )\left (
26\,t^{2}+46\,tu+26\,u^{2}
\right )\ln (-t/-u)}\over{s^{5}}}
\biggr)
}
\eqn\FourPhotSUGRA
$$
where $F'$ here is
$$
{i\kappa^4(4\pi)^{\eps}r_{\Gamma}\over 16(4\pi)^2}
 {\spa4.3^2 \over \spa1.2^2}\biggl( {st \spa1.2^4 \over
\spa1.2\spa2.3\spa3.4\spa4.1 } \biggr)^2 =
{stu\kappa^2(4\pi)^{\eps}r_{\Gamma}\over 4(4\pi)^2}
{\Atree(\gamma^-,\gamma^-,\gamma^+,\gamma^+)}.
\anoneqn
$$
 It is clear that the divergences seen here are the complete IR
contributions expected; there are no UV contributions from this `$N=2$'
part of the calculation.  So, the UV divergence we are looking for will be
found in the subtracted gravitino contribution. 
  
\goodbreak
 
In fact, we only need to carry out one calculation to find the divergences
in the gravitino contribution: Note that if the helicities on one side of
the cut are the same then at least one tree will vanish.  Also, there is a
symmetry between the two non-zero cut contributions -- those in the $t$-
and $u$-channels.  So, if we calculate the $t$-channels divergences, we
will have the complete result. Remarkably, using a combination of SUSY and 
previous results, we have been able to reduced the work required to a single 
calculation. 

Let us look at the $t$-channel contribution. For this we will need the
photon-gravitino tree. This can be found via another supersymmetric relation:
$$\eqalign{ A^{\rm N=2}(\gamma^{-} \psi^{-} \gamma^{+} \psi^{+}) &=
{\spa3.2\spa3.4  \over \spa1.2^2}
    A^{\rm N=2}(g^{-} g^{-} g^{+} g^{+}). }
\anoneqn
$$
Using this for a cut calculation gives us the result
$$
{i\kappa^4{r_\Gamma}\over(4\pi)^{2-\eps}}
{s^4\over\spa2.1^2\spb3.4^2}
\left({137\over960}\ln(t)+{137\over960}\ln(u)-\ln^2(t/u)\right).
\anoneqn
$$
 We must double this since there are two gravitinos in the SUSY multiplet;
the contributions from both should be subtracted. So, we can deduce that the
UV divergence is
 $$
{137\over240}{i\kappa^4\over(4\pi)^{2}}{1\over\eps}{s^4\over\spa2.1^2\spb3.4^2}.
\anoneqn
$$
 We can compare this with the theoretical
derivation of the counterterms (see
ref.~[\use\EMGrav]\footnote{$^\dagger$}{This actually differs from the
result in ref.~[\use\EMGrav] by a factor of 4.  This is simply due to a
difference between their definition of $\kappa$ and the definition
implicit in our calculation.}); we confirm that these terms do,
indeed, cancel this, as required. 
 
\midinsert{\medskip
\centerline{\epsfxsize=3in \epsfbox{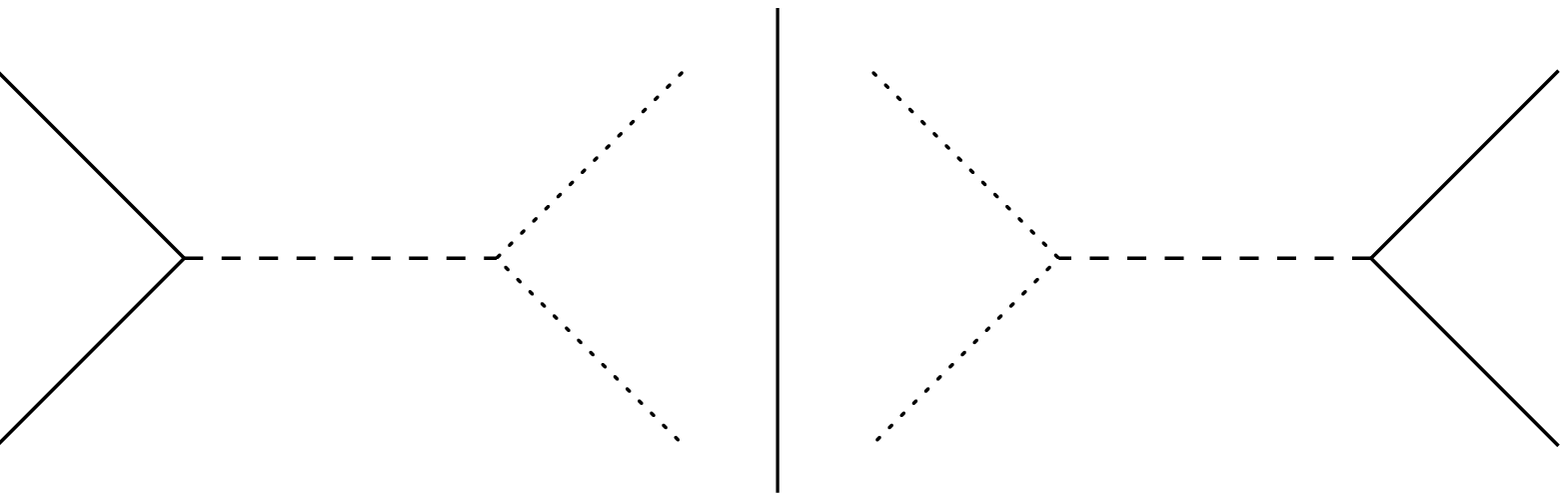}}
\smallskip
{\baselineskip 10 pt\ninerm
\centerline{\fign\PhotH\ Extra cut 
required 
when a second photon flavour} \centerline{is added to the theory.}
}}
\endinsert

\cut
We can also look at the effect of adding more (independent) $U(1)$
particles to the system.  To add $n$ more photons, we must consider $n$
diagrams of the form fig.~\use\PhotH, where the internal particle is a
photon, but a different flavour to the external one. 
If we carry out such a calculation, and sum over all channels,
we find that $n$ extra photons produce an added UV infinite contribution
of
$${n\over40} 
{i\kappa^4\over(4\pi)^{2}}{1\over\eps}{s^4\over \spa2.1^2\spb3.4^2}.
\eqn\ExtraPh
$$
Again, comparing this with the theoretical results~[\use\YMGrav], we
find that the derived counterterms will remove this divergence.

\cut In conclusion, we have shown how the Cutkosky rules can be used to
find infinities in two quantum gravity theories.  In the
Dirac-Einstein case this enabled us to determine an unknown coefficient in
the counter-Lagrangian. We were also able to confirm the previous
derivation of Einstein-Maxwell counterterms. In the latter calculation we 
found that SUSY Ward identities could simplify the process significantly.

\cut
 It is a pleasure to thank Dave Dunbar and Graham Shore for useful
conversations.  This work was funded by P.P.A.R.C. 

\listrefs

\end